

   \magnification\magstep1 
   \baselineskip = 0.6 true cm
                           
  \def\sa{\vskip 0.30 true cm}
  \def\sb{\vskip 0.60 true cm}

  \nopagenumbers
  \voffset = 15 true mm
  \hoffset =  8 true mm
  \hsize = 15 true cm
  \vsize = 22 true cm



\def\souligne#1{$\underline{\smash{ \hbox{#1}}}$}

\rightline{\bf LYCEN 9145}
\rightline{November 1991}

\sa
\sb
\sa

\centerline {{\bf SYMMETRY ADAPTATION IN TWO-PHOTON SPECTROSCOPY}}

\sa
\sb

\centerline {Maurice Kibler}

\sa

\centerline {Institut de Physique Nucl\'eaire de Lyon} 
\centerline {IN2P3-CNRS et Universit\'e Claude Bernard} 
\centerline {43 Boulevard du 11 Novembre 1918} 
\centerline {F-69622 Villeurbanne Cedex, France} 

\sa
\sa
\sb
\sb
\sb

\baselineskip = 0.55 true cm

\sa

\centerline {\bf ABSTRACT}

\sa
\sb

\noindent Symmetry adaptation techniques are applied to the 
determination of the intensity of two-photon 
transitions for transition ions in finite symmetry 
environments. We treat the case of intra-configurational 
transitions with some details and briefly report some results 
on inter-configurational transitions. In particular, for 
intra-configurational transitions, we describe a model which 
takes into account the following ingredients~: (symmetry, 
second- plus third-order mechanisms, $S$-, $L$- and $J$-mixings).

\sa
\sb
\sa
\sb
\sa
\sb
\sb
\sb

\noindent Invited lecture at the "Second International 
School on Excited States of Transition Elements", 
Wroc\l aw-Karpacz, Poland, 2~-~6 September 1991. 
Published in {\bf Excited States of Transition Elements,
Eds. W. Str\c ek, W. Ryba-Romanowski, J. Legendziewicz, and B.
Jez\.owska-Trzebiatowska (World Scientific, Singapore, 1992),
p.~139-148}.



\vfill\eject
\baselineskip = 0.5 true cm

\centerline {{\bf SYMMETRY ADAPTATION IN TWO-PHOTON SPECTROSCOPY}}

  \sb

\centerline {Maurice Kibler} 

\sa

\centerline {Institut de Physique Nucl\'eaire de Lyon}
\centerline {IN2P3-CNRS et Universit\'e Claude Bernard}
\centerline {F-69622 Villeurbanne Cedex, France} 
  \sa
  \sb
\baselineskip = 0.5 true cm
\centerline {ABSTRACT}

\leftskip  = 1.4 true cm
\rightskip = 1.4 true cm

\noindent Symmetry adaptation techniques are applied to the 
determination of the intensity 
of two-photon (intra- and inter-configurational) 
transitions for transition ions in finite symmetry 
environments. 

\leftskip  = 0 true cm
\rightskip = 0 true cm

  \sa
  \sb
\baselineskip = 0.5 true cm
\noindent {1. INTRODUCTION}

Symmetry adaptation techniques, developed in the spirit of 
Refs.~[1,2], for a chain of groups $O(3) \supset G$ are applied here 
to two-photon spectroscopy of transition ions of 
configuration $n \ell^N$ ($\ell = d$ for transition metal ions 
                      and $\ell = f$ for lanthanide or 
actinide ions) in surroundings of finite symmetry 
$G$. More precisely, we show in this lecture 
how Wigner-Racah calculus 
for a chain $O(3) \supset G$ (in terms of simple or 
double groups) can be combined with models$^{3-20)}$ based on second- 
plus third-order mechanisms in order to isolate the 
polarization dependence from the 
intensity of two-photon transitions for an $n \ell^N$ ion 
in a molecular or solid-state environment 
with symmetry $G$. (For classification and 
symmetry-breaking purposes, the group $G$ may be replaced by a 
chain of subgroups of $O(3)$, the relevant symmetry group being 
one of the groups of the chain.)  

The r\^ole of symmetries in two-photon spectroscopy of 
partly-filled shell ions in finite symmetry 
is touched upon in Refs.~[4-6,12-14,16,17]. In 
Refs.~[4,5], the information arising from 
symmetry is handled mainly in a qualitative way. 
More quantitative results can be found in Refs.~[6,12-14,16,17]. 
In Ref.~[12], the accent is put on the transition 
matrix elements between initial and final state vectors while 
emphasis is on the intensity strength in Refs.~[14,16] and in the 
present lecture. This lecture constitutes a complement to the 
material presented in Refs.~[14,16]. 

Two distinct cases are studied in this work. The case of 
($n \ell^N \to n \ell^N$, e.g., $3d^N \to 3d^N$
                            and $4f^N \to 4f^N$) 
intra-configurational two-photon 
transitions, which are parity allowed, is worked out in section 
2 and the one of 
($n \ell^N \to n \ell^{N-1} n' \ell'$ with $\ell + \ell'$ odd, 
e.g., $3d^N \to 3d^{N-1}4p$ 
  and $4f^N \to 4f^{N-1}5d$) 
inter-configurational two-photon transitions, which are parity forbidden, is 
examined in section 3.  

\noindent {2. INTRA-CONFIGURATIONAL TWO-PHOTON TRANSITIONS}

\noindent {2.1. Preliminaries}

We know that the electronic transition matrix element $M_{i \to f}$ between an
initial state $i$ and a final state $f$ is, in the framework of the 
electric dipolar approximation, given by
$$
M_{i \to f} = \sum_v {{1}\over {\Delta_1}} 
\left(f \vert \vec D. \, \vec {\cal E}_2 \vert v \right) 
\left(v \vert \vec D. \, \vec {\cal E}_1 \vert i \right) 
            + \sum_v {{1}\over {\Delta_2}} 
\left(f \vert \vec D. \, \vec {\cal E}_1 \vert v \right) 
\left(v \vert \vec D. \, \vec {\cal E}_2 \vert i \right) 
\eqno (1)
$$
The two summations in (1) have to be extended over all the 
intermediate states $v$ having a parity different 
from the one of the states $i$ and $f$. Furthermore, we have 
$\Delta_\lambda = \hbar \omega_{\lambda} - E_v$, 
where $E_v$ is the energy of the state $v$ with respect to 
that of the state $i$ and
$\hbar \omega_{\lambda}$ the energy of the photon no.~$\lambda$. 
(For Raman scattering, the sign of $\hbar \omega_2$ has to be 
changed.) The quantity 
$\vec D . \, {\vec {\cal E}}_{\lambda}$ in (1) stands for 
the scalar product of the electric dipolar moment operator 
$\vec D$ for the $N$ electrons and the unit polarization vector 
${\vec {\cal E}}_\lambda$ for the photon no.~$\lambda$. (We use 
single-mode excitations, of energy $\hbar \omega_{\lambda}$, 
                    wave-vector ${\vec k}_{\lambda}$ and 
                   polarization ${\vec {\cal E}}_{\lambda}$, 
for the radiation field.) The two photon beams 
can be polarized either circularly with
$$
({\cal E}_\lambda)_q = - \delta (q,-1) 
\quad \hbox {if} \quad {\vec {\cal E}}_\lambda = {\vec e}_{+1} 
\quad \hbox {while} \quad 
({\cal E}_\lambda)_q = - \delta (q,+1) 
\quad \hbox {if} \quad {\vec {\cal E}}_\lambda = {\vec e}_{-1} 
\eqno (2)
$$
or linearly with
$$
({\cal E}_\lambda)_0 = \cos \theta_\lambda \qquad 
({\cal E}_\lambda)_{\pm 1} = \mp
{1 \over {\sqrt 2}} \; \sin \theta_\lambda \; \exp (\pm i \varphi_\lambda) 
\eqno (3)
$$
In equations (2) and (3), we use the components 
$({\cal E}_\lambda)_q = {\vec {\cal E}}_\lambda . \, {\vec e}_q$ 
(with $q = - 1, 0, 1$ and $\lambda = 1,2$) 
in the standard spherical basis 
$\left(\vec e_{-1}, \vec e_0, \vec e_{+1}\right)$.
In the case of a linear polarization, the angles 
$(\theta_\lambda, {\varphi}_\lambda)$ 
are the polar angles of the polarization vector 
${\vec {\cal E}}_\lambda$ 
($\lambda = 1,2$) with respect to the crystallographic $c$-axis. 
For two-photon absorption, only one sum occurs in (1) when the 
two photons are identical. 

Equation (1) can be derived from the time-dependent perturbation theory 
and goes back to the work of G\"oppert-Mayer$^{18,19)}$ (see the 
lecture by J.C. G\^acon in these proceedings). It
is also possible to derive it, in an elegant way, from the method of the
resolvent operator$^{20)}$.

\noindent {2.2. State vectors}

The initial state $i$ with symmetry $\Gamma$ is characterized by the state
vectors $\vert i \Gamma \gamma)$ where $\gamma$ 
($\gamma = 1$, 2, $\cdots$, dim~$\Gamma$) is a
multiplicity label to be used if the dimension dim~$\Gamma$ of the irreducible
representation class (IRC) $\Gamma$ of the group $G$ is greater than 1. The
state vector $\vert i \Gamma \gamma)$ is taken in the form
$$
\vert i \Gamma \gamma) \equiv \vert n \ell^N  i \Gamma \gamma) = 
\sum_{\alpha S L J a} \; \vert n \ell^N \alpha S L J a \Gamma \gamma) 
\; c(\alpha S L J a \Gamma; i) \eqno (4)
$$
in terms of the $O(3) \supset G$ symmetry adapted state 
vectors$^{1,2)}$
$$
\vert n \ell^N \alpha S L J a \Gamma \gamma) \; = \; \sum_{M=-J}^J \; 
\vert n \ell^N \alpha S L J M) \;  (J M \vert J a \Gamma \gamma) 
\eqno (5)
$$
The coefficients $(JM \vert J a \Gamma \gamma)$ in (5) are reduction
coefficients to pass from               the chain $O(3) \supset O(2)$ 
characterizing the $\{ JM \}$ scheme to the chain $O(3) \supset G$ 
characterizing the $\{ J a \Gamma \gamma \}$ scheme ; they 
depend on the group $G$ with a certain degree of freedom 
emphasized by the branching multiplicity label $a$ to be used when $\Gamma$
occurs several times in the IRC $(J)$ of $O(3)$. 
In contradistinction, the coefficients $c(\alpha S L J a \Gamma ; i)$ 
in (4) depend on the Hamiltonian employed for obtaining
the initial state $i$. Similarly, for the final state $f$ with symmetry 
$\Gamma'$, we have the state vectors 
$$
\vert f \Gamma' \gamma') \; \equiv \; \vert n \ell^N f \Gamma' \gamma') 
\; = \; \sum_{\alpha' S' L' J' a'} \; \vert n \ell^N \alpha' S' L' J' a' \Gamma'
\gamma') \; c (\alpha' S' L' J' a' \Gamma' ; f) \eqno (6)
$$
in terms of $O(3) \supset G$ symmetry adapted state vectors. 
The only good quantum numbers for the initial and final state vectors are
$\Gamma \gamma$ and $\Gamma' \gamma'$, respectively. Although, the state
vectors $\vert i \Gamma  \gamma )$ and $\vert f \Gamma' \gamma')$ 
are developed in a weak-field basis, it is to be noted that
the intensity calculation to be conducted in what follows is valid for any 
strength (weak, intermediate or strong) of the crystalline 
field. 

\noindent {2.3. Transition matrix element}

By using a quasi-closure approximation, it can be shown that the transition 
matrix element $M_{i \to f}$ between the state vectors 
$\vert i \Gamma  \gamma )$ and 
$\vert f \Gamma' \gamma')$ is given by 
$$
M_{i \to f} \; \equiv \; M_{i (\Gamma \gamma) \to f (\Gamma' \gamma')} 
\; = \; (f \Gamma' \gamma' \vert H_{eff} \vert i \Gamma \gamma) \eqno (7)
$$
where $H_{eff}$ is an effective operator$^{7,8)}$. This operator may 
be written as$^{12)}$ 
$$
H_{eff} \; = \; \sum_{k=0,1,2} \; \sum_{k_Sk_L} 
  \; C \left[ \left( k_S k_L \right) k \right] 
  \; \left( \left\{ {\cal E}_1 \, {\cal E}_2 \right\} ^{(k)} . \, 
  \; {\bf W}^{(k_Sk_L)k} \right) \eqno (8)
$$
In equation (8), ${\bf W}^{(k_Sk_L)k}$ is an electronic double tensor of 
spin rank $k_S$, orbital rank $k_L$ and total rank $k$. The information on the
polarization of the two photons is contained in the tensor product
$\left\{ {\cal E}_1 \, {\cal E}_2 \right\} ^{(k)}$ 
of rank $k = 0$, 1 or 2. The right-hand side
of (8) is a development in terms of scalar products $(\,.\,)$ with expansion
coefficients $C\left[\left(k_S k_L\right)k\right]$. These coefficients depend
on the ground configuration $n \ell^N$ and on the configurations 
$n  \ell ^{N-1} n' \ell'$ and/or 
$n' \ell'^{4 \ell' + 1} n \ell^{N +1}$, with $\ell + \ell'$ odd, 
from which the states $v$ arise. 

Only the contributions 
$(k_S = 0, k_L = 1, k = 1)$ and 
$(k_S = 0, k_L = 2, k = 2)$ correspond to the
standard theory originally developed by Axe$^{3)}$. The other contributions 
$(k_S \ne 0, k_L, k)$, which may include 
$(k_S = 1, k_L = 1, k = 0)$ and 
$(k_S = 1, k_L = 1, k = 2)$, 
correspond either to mechanisms introduced by various 
authors$^{7-11)}$ or to phenomenological contributions$^{12)}$. 
The contributions $(k_S =   0, k_L = k, k)$ and 
                                $(k_S \ne 0, k_L    , k)$ are often referred 
to as second-order and third-order mechanisms, respectively. It is in principle
possible to find an expression for the parameters 
$C \left[ \left( k_S k_L \right) k \right] $. 
Among the various 
contributions $(k_S \ne 0, k_L, k)$, the contribution 
$(k_S = 1, k_L = 1, k = 0)$ arises from the spin-orbit 
interaction within the configuration $n \ell^{N-1} n' \ell'$ as 
was shown for lanthanide ions$^{7,8)}$.

The transition matrix element (7) is easily calculated by means of 
Wigner-Racah calculus for the chain $O(3) \supset G$. As a result, 
we have$^{12)}$ 
$$
\eqalign{
M_{i(\Gamma \gamma) \to f (\Gamma' \gamma')} = & \sum_{\alpha' S' L' J' a'} \; 
                                                 \sum_{\alpha  S  L  J  a } 
\; c (\alpha' S' L' J' a' \Gamma' ; f)^* 
\; c (\alpha  S  L  J  a  \Gamma  ; i) \cr 
\sum_{k_S k_L k} \; (-)^{k_S + k_L - k} & \; 
C \left[ \left( k_S k_L \right) k\right] \; 
\left( n \ell^N \alpha  S  L  J \Vert W^{(k_S k_L)k} \Vert 
       n \ell^N \alpha' S' L' J' \right) ^* \cr
\sum_{a'' \Gamma'' \gamma''} & \; f
\pmatrix{
J&J'&k\cr
a \Gamma \gamma& a' \Gamma' \gamma'& a'' \Gamma'' \gamma''\cr
}^* \,  
\left\{ {\cal E}_1 \, {\cal E}_2 \right\} ^{(k)}_{a'' \Gamma'' \gamma''}
}
\eqno (9)
$$
where the $f$ symbol denotes an $O(3) \supset G$ symmetry adapted coupling 
coefficient defined by$^{1,2)}$ 
$$
\eqalign{
f
\pmatrix{
J&J'&k\cr
a \Gamma \gamma & a' \Gamma' \gamma' & a'' \Gamma'' \gamma''\cr
}
\; = \; & \sum_{M M' q} \; (-)^{J-M} 
\pmatrix{
J&k&J'\cr
-M&q&M'\cr
} \cr
& (J  M  | J  a   \Gamma   \gamma )^* \; 
  (J' M' | J' a'  \Gamma'  \gamma')   \; 
  (k  q  |  k a'' \Gamma'' \gamma'') 
} 
$$
Equation (9) follows by developing (7) 
with the help of (4), (6) and (8).

\noindent {2.4. Intensity formula} 

The quantity of interest for a comparison between theory and experiment is the
intensity $S_{i(\Gamma) \to f(\Gamma')}$ of the two-photon transition between
the initial state $i$ and the final state $f$. This intensity is given by
$$
S_{  \Gamma  \to   \Gamma' } \; \equiv \; 
S_{i(\Gamma) \to f(\Gamma')} \; = \; \sum_{\gamma \gamma'} \; 
\left\vert M_{i(\Gamma \gamma) \to f(\Gamma' \gamma')} \right\vert ^2 
\eqno (10)
$$
By introducing (9) into (10) and using the factorization 
property$^{1)}$ for the $f$ coefficients as well as the 
orthonormality-completeness property$^{1)}$ for the Clebsch-Gordan 
coefficients of the group $G$, we obtain the compact expression
$$
S_{\Gamma \to \Gamma'} \; = \; \sum_{k,\ell} \; \sum_{r,s} \; \sum_{\Gamma''} \;
I [k \ell r s \Gamma'' ; \Gamma \Gamma'] \; \sum_{\gamma''} \; \left\{ 
{\cal E}_1 \, {\cal E}_2 \right\} ^{(k)}    _{r \Gamma'' \gamma''} \; 
\left( \left\{ 
{\cal E}_1 \, {\cal E}_2 \right\} ^{(\ell)} _{s \Gamma'' \gamma''}\right)^* 
\eqno (11)
$$
In equation (11), the parameter $I$ reads
$$
\eqalign{
I [k \ell & r s \Gamma'' ; \Gamma \Gamma'] \; = \; [\Gamma'']^{-1} \; [\Gamma] 
\; \sum_{J'a'} \; 
   \sum_{J a } \; 
\sum_{\bar J' \bar a'} \; 
\sum_{\bar J  \bar a}  \; 
  Y_k    (     J'      a' \Gamma' ,      J      a \Gamma) \cr 
& Y_\ell (\bar J' \bar a' \Gamma' , \bar J \bar a \Gamma)^* \; 
\sum_\beta \; (J' a' \Gamma' + k r \Gamma'' \vert J a \beta \Gamma) \; 
(\bar J' \bar a' \Gamma'+\ell s \Gamma'' \vert \bar J \bar a \beta \Gamma)^* 
}
\eqno (12)
$$
where $Y_k$ is defined by
$$
\eqalign{
Y_k (J' a' \Gamma', & J a \Gamma) = [J]^{-1/2} \sum_{\alpha' S' L'} 
\sum_{\alpha S L} \sum_{k_S k_L} 
c (\alpha' S' L' J' a' \Gamma' ; f)^* \; 
c (\alpha  S  L  J  a  \Gamma  ; i) \cr 
& C [(k_S k_L) k] \; (-)^{k_S + k_L - k} \; 
(n \ell^N \alpha S L J \Vert W^{(k_Sk_L)k} \Vert n \ell^N \alpha' S' L' J')^* 
}
\eqno (13)
$$
and $Y_\ell$ by a relation similar to (13). In (12) the 
$( \; + \; \vert \; )$ coefficients stand for isoscalar factors 
of the chain $O(3) \supset G$ and the labels $\beta$ are 
internal multiplicity labels to be used for those Kronecker products 
which are not multiplicity-free$^{1)}$. 

\noindent {2.5. Properties and rules}

The $I$ parameters in (11) can be calculated in an {\it ab initio} way or 
can be considered as phenomenological parameters. In both approaches, the 
following properties and rules are of central importance.

{Property 1}. In the general case, we have the (hermitean) property
$$
I[\ell k s r \Gamma'' ; \Gamma \Gamma']^* \; = \; 
I[k \ell r s \Gamma'' ; \Gamma \Gamma'] \eqno (14)
$$
which ensures that $S_{\Gamma \to \Gamma'}$ is a real quantity.

{Property 2}. In the case where the group $G$ is multiplicity-free, 
we have the factorization formula
$$
I[k \ell r s \Gamma'' ; \Gamma \Gamma'] \; = \; 
\chi [k    r \Gamma'' ; \Gamma \Gamma'] \; 
\chi [\ell s \Gamma'' ; \Gamma \Gamma']^*
\eqno (15)
$$
where the function $\chi$ is defined through
$$
\chi [k    r \Gamma'' ; \Gamma \Gamma'] \; = \; 
[\Gamma'']^{-1/2} \; [\Gamma]^{1/2} \; \sum_{J'a'} \; \sum_{Ja} \; 
Y_k(J' a' \Gamma', J a \Gamma) \; 
(J' a' \Gamma' + k r \Gamma'' \vert J a \Gamma)
$$
(In a less restrictive sense, equation (15) is valid when the Kronecker 
product $\Gamma'^* \otimes \Gamma$, of the complex conjugate IRC of 
$\Gamma'$ by the IRC $\Gamma$, is multiplicity-free.)

The number of independent parameters $I$ in the expansion 
(11) can be {\it a priori} determined from the two following selection 
rules used in conjunction with Properties 1 and 2.

{Rule 1}. In order to have $S_{\Gamma \to \Gamma'} \ne 0$, it is
necessary that
$$
\Gamma'' \subset \Gamma'^* \otimes \Gamma \qquad \quad 
\Gamma'' \subset (k_g)                    \qquad \quad 
\Gamma'' \subset (\ell_g) 
\eqno (16)
$$
where $(k_g)$ and $(\ell_g)$ are {\it gerade} IRC's of the group 
$O(3)$ associated to the integers $k$ and $\ell$, respectively.

{Rule 2}. The sum over $k$ and $\ell$ in the intensity 
formula (11) is partially controlled by the selection rule
$$
 {\cal E}_1 \ne {\cal E}_2: \ 
k, \ell = 1,2     \ \hbox {for 2nd-order} \quad \hbox {or}  \quad 
k, \ell = 0, 1, 2 \ \hbox {for 2nd-order} + \hbox {3rd-order} 
$$
or 
$$
 {\cal E}_1 =   {\cal E}_2: \
k, \ell = 2       \ \hbox {for 2nd-order} \quad \hbox {or}  \quad 
k, \ell = 0, 2    \ \hbox {for 2nd-order} + \hbox {3rd-order} 
$$
according to as the two photons have different or the same polarization. 
(Note that the situation ${\cal {E}}_1 = {\cal {E}}_2$ surely occurs for 
identical photons but may also occur for non-identical photons.)

\noindent {2.6. Discussion}

For most of the cases of interest, there is no summation on $r$ and $s$, two
branching multiplicity labels of type $a$, in the intensity formula 
(11). (In other words, the frequency of $\Gamma''$ in $(k_g)$ and $(\ell_g)$ 
is rarely greater than 1.) The group-theoretical selection rules 
(16) impose strong limitations
on the summation over $\Gamma''$ in (11) once $\Gamma$ and $\Gamma'$ are
fixed and the range of values of $k$ and $\ell$ is chosen.

The number of independent intensity parameters $I$ in 
the formula (11) depends on~: (i) the nature of the photons, cf.~Rule 2~;
(ii) the group $G$, cf.~Rule 1~; (iii) the conjugation 
property (14), cf.~Property 
1~; (iv) the use of $k_S = 0$ (second-order mechanisms) or $k_S = 0$ and
$k_S \not = 0$ (second- plus third-order mechanisms), cf.~Rule 2~; 
(v) the (weak-, intermediate- or strong-field) state vectors 
used in conjunction with equations (12) and (13).

Points (i)-(iii) depend on external physical conditions. On the other hand,
points (iv) and (v) are model-dependent. In particular, in the case where the
$J$-mixing, cf.~point (v), can be neglected, a situation often of interest for
lanthanide ions, the summations on $k$ and $\ell$ in (11) are further reduced
by the triangular rule $|J-J'| \leq k, \ell \leq J+J'$, where $J$ and $J'$ are
the total angular quantum numbers for the initial and final states,
respectively. Similar restrictions apply to $k_S$ and $k_L$ in 
(13) if the $S$- and $L$-mixing are neglected.

The computation, via equations (12) and (13), of the $I$ parameters generally 
is a difficult task. Therefore, they may be considered, at least in a
first step, as phenomenological parameters. In this respect, equations 
(12) and (13) should serve as a guide for reducing the number of $I$ parameters.

Once the number of independent parameters $I$ in the 
formula (11) has been determined, we can obtain the polarization dependence 
of the intensity strength 
$S_{\Gamma \to \Gamma'}$ by calculating the tensor products 
$\left\{ {\cal E}_1 \, {\cal E}_2 \right\} ^{(K)} _{a'' \Gamma'' \gamma''}$ 
(with $K = k, \ell$ and $a'' = r, s$) occurring in (11). 
For this purpose, we use the development 
$$
\left\{ {\cal E}_1 \, {\cal E}_2 \right\} ^{(K)} _{a''\Gamma'' \gamma''} 
\; = \;  \sum^K_{Q = - K} \;
\left\{ {\cal E}_1 \, {\cal E}_2 \right\} ^{(K)} _Q \; 
(K Q \vert K a'' \Gamma'' \gamma'')
\eqno (17)
$$
in terms of the spherical components 
$\left\{ {\cal E}_1 \, {\cal E}_2 \right\} ^{(K)} _Q$,
the coefficients in (17) being reduction coefficients 
for the chain $O(3) \supset G$. Then, we use in turn the development
$$
\left\{ {\cal E}_1 \, {\cal E}_2 \right\} ^{(K)} _Q \; = \; 
(-)^{K - Q} \; [K]^{1/2} \; \sum^1_{m = - 1} \; \sum^1_{m' = -1} \; 
\pmatrix{
1&K&1\cr
m&-Q&m'\cr
}
\; ({\cal E}_1)_m 
\; ({\cal E}_2)_{m'} 
$$
in terms of the spherical components $({\cal E}_\lambda)_q$ defined 
by (2) or (3) for circular or linear polarization, respectively.

\noindent {2.7. Illustration}

As a pedagological example, let us consider the case of the 
two-photon absorption (intra-configurational) 
transition $^7F_0$ $\to$ $^5D_0$ for the 
configuration $4f^6$ in tetragonal symmetry 
(with $G \equiv C_{4v}$ or $D_{2d}$). In this case, we have 
$\Gamma = A_1$ for the initial state and $\Gamma ' = A_1$ for 
the final state, whatever the strength of the crystalline field 
is. Rule 1 then yields $\Gamma'' = A_1$ and, consequently, 
there is no sum on the label $\gamma''$ in (11). Furthermore, 
there is no sum on the branching multiplicity labels $r$ and $s$ 
in the intensity formula (11). Let us consider an experimental 
situation where the two photons are identical (one-color beam 
arrangement) so that 
${\cal E}_1 \equiv 
 {\cal E}_2 = 
 {\cal E}$. We continue with a model characterized (in 
addition to the symmetry $C_{4v}$ or $D_{2d}$) by the use of 
second- plus third-order mechanisms. Thus, according to Rule 2, 
the indices $k,\ell$ in (11) may assume the values $0$ and $2$. 
By introducing the abbreviation 
$I( k \ell ) \; \equiv \; 
I[k \ell r s \Gamma'' = A_1 ; \Gamma = A_1 \ \Gamma' = A_1]$, 
we are left with 3 {\it a priori} independent parameters (cf. 
Property 1), viz.,
$I(00)$, $I(02) = I(20)^*$ and $I(22)$. 
Taking the wave-vector $\vec k$ of the two photons parallel to 
the crystallographic $c$-axis, we have
$$
  \{ {\cal E} {\cal E} \}^0_{A_1} \equiv 
  \{ {\cal E} {\cal E} \}^0_{0}     = {-1 \over \sqrt{3}} 
  \ {\rm or} \ 0 \quad \hbox {and} \quad
  \{ {\cal E} {\cal E} \}^2_{A_1} \equiv 
  \{ {\cal E} {\cal E} \}^2_{0}     = 
  {{3 \cos^2 \theta - 1} \over \sqrt{6}} 
  \ {\rm or} \ 0 
  \eqno (18)
$$
for linear or circular polarization, respectively. Finally, the 
intensity strength $S_{A_1 \to A_1}$ is easily obtained by 
introducing (18) into (11). This leads to 
$$
S_{A_1 \to A_1} \; = \; r^2 \; + \; s   (3 \cos^2 \theta - 1)
                            \; + \; t^2 (3 \cos^2 \theta - 1)^2 
\qquad {\rm or} \qquad 0
\eqno (19)
$$ 
according to whether as the polarization is linear or circular, 
respectively. Here, the real parameter $s$ and the 
two non-negative parameters $r^2$ and $t^2$ are defined by
$$
r^2 = {I(00) \over 3} \quad \qquad 
s = {{ - I(02) - I(02)^*} \over 3 \sqrt{2}} \quad \qquad 
t^2 = {I(22) \over 6}
$$
In the intensity formula (19), the two first terms (in $r^2$ and 
$s$) arise from third-order mechanisms while the third one (in 
$t^2$) comes from second-order mechanisms. Only the scalar 
term (in $r^2$) contributes to $S_{A_1 \to A_1}$ 
in the absence of $J$-mixing. 

Equation (19) has been applied$^{13)}$ to the cases of 
Sm$^{2+}$:BaClF, Sm$^{2+}$:SrClF and 
Eu$^{3+}$:LuPO$_4$. Neither the conventional 
second-order term (in $t^2$) nor the scalar third-order term 
(in $r^2$) are sufficient to reproduce the experimental data. 
Indeed, on the basis of fitting procedures$^{13)}$, a good agreement 
between theory and experiment requires in these cases that the 
three terms (in $r^2$, $t$, and $s^2$) contribute to the 
intensity strength (19). As a conclusion, the model inherent to 
equation (19), the ingredients of this model being~: (symmetry, second- 
plus third-order mechanisms, $J$-mixing), is appropriate to the 
$^7F_0$ $\to$ $^5D_0$ two-photon transition for 
Sm$^{2+}$:BaClF, Sm$^{2+}$:SrClF and Eu$^{3+}$:LuPO$_4$. 

Similar models have been applied to the two-photon 
transitions$^{13)}$ $^7F_0(\Gamma=A_1)$ 
$\to$ $^5D_2(\Gamma'=A_1,B_1,B_2,E)$ for the tetragonal 
compound Sm$^{2+}$:BaClF and to 
the two-photon transitions$^{17)}$ 
$^3A_2(\Gamma=T_2)$ $\to$ $^3T_2(\Gamma'=A_2,E,T_1,T_2)$ 
for the cubical compound Ni$^{2+}$:MgO. In both cases, we have 
found that second-order mechanisms are sufficient to describe 
the transitions. More precisely, the model~: (symmetry, 
second-order mechanisms, $S$- and $L$-mixing but no $J$-mixing) 
works for Sm$^{2+}$:BaClF 
while the model~: (symmetry, second-order mechanisms, 
$S$-, $L$- and $J$-mixing) works for Ni$^{2+}$:MgO. 

\noindent {3. INTER-CONFIGURATIONAL TWO-PHOTON TRANSITIONS}

\noindent {3.1. Sketch of the theory}

We now consider two-photon transitions between Stark levels 
arising from the configurations $n \ell^N$ and 
$n \ell^{N-1} n' \ell'$ of opposite parities 
($\ell + \ell'$ odd). For the sake of simplicity, we 
deal here with identical photons. The initial 
(i.e., $\vert i \Gamma  \gamma )$)
and final 
(i.e., $\vert f \Gamma' \gamma')$)
state vectors are taken in the form 
$$
\eqalign{
\vert n \ell^N  i \Gamma \gamma) = 
\sum_{\alpha S L J a} \; & \vert n \ell^N \alpha S L J a \Gamma \gamma) 
\; c (\alpha  S  L  J  a  \Gamma  ; i) \cr
\vert n \ell^{N-1} n' \ell' f \Gamma' \gamma') = 
\sum_{\alpha' S' L' J' a'} \; 
& \vert n \ell^{N-1} n' \ell'   \alpha' S' L' J' a' \Gamma' \gamma') 
\; c (\alpha' S' L' J' a' \Gamma' ; f) 
} \eqno (20)
$$
to be compared with equations (4) and (6).

It is clear that the transition matrix element
$$
M_{i(\Gamma \gamma) \to f(\Gamma' \gamma')} \; = \; \sum_v \; 
{{1}\over {\Delta}} \; 
\left( f \Gamma' \gamma'  \vert \vec D. \, \vec {\cal E} \vert 
      v \Gamma_v \gamma_v \right) 
\left(v \Gamma_v \gamma_v \vert \vec D. \, \vec {\cal E} 
\vert i \Gamma \gamma \right) 
$$
is identically zero. In order to obtain 
$M_{i(\Gamma \gamma) \to f(\Gamma' \gamma')} \ne 0$, it is 
necessary to pollute (20), as well as the intermediate state 
vectors, with state 
vectors of the type $\vert n \ell^{N-1} n' \ell' x' \Gamma' \gamma')$ 
and                 $\vert n \ell^{N}            x  \Gamma  \gamma )$, 
respectively. This may be 
achieved by using first-order time-independent perturbation theory where the 
polluting agent is the crystal-field potential $H_3$ of odd order, 
which is static or dynamic according to as the group $G$ does not or does 
have a center of inversion. Hence, we produce state vectors 
noted $\vert n \ell^ N             i \Gamma  \gamma >$ and 
      $\vert n \ell^{N-1} n' \ell' f \Gamma' \gamma'>$ from which we 
can calculate, in a 2nd-order time-dependent plus 1st-order 
time-independent scheme, a non-vanishing transition matrix element 
$$
M_{i(\Gamma \gamma) \to f(\Gamma' \gamma')} \; = \; \sum_v \; 
{{1}\over {\Delta}} \; <f \Gamma' \gamma' 
                     \vert \vec D. \, \vec {\cal E} \vert v \Gamma_v \gamma_v>
<v \Gamma_v \gamma_v \vert \vec D. \, \vec {\cal E} \vert i \Gamma \gamma> 
$$
Then, we apply a quasi-closure approximation both for the 
initial, intermediate, and final state 
vectors and the transition matrix element. This approximation can be 
summarized by 
$E(n' \ell') - E(n \ell) \; = \; 2 \, \hbar \, \omega$. 
We thus obtain a closed form formula for 
$M_{i(\Gamma \gamma) \to f(\Gamma' \gamma')}$ (see Ref.~[20]). 

At this stage, it should be mentioned that the so-obtained 
formula is equivalent to that we would obtain, within the 
just mentioned approximation, by using third-order mechanisms described by
$$
\eqalign{
M_{i(\Gamma \gamma) \to f(\Gamma' \gamma')} & = 
\sum_{v_1 v_2} \; {{1}\over {\Delta(v_1)}} \; {{1}\over {\Delta(v_2)}} 
\; ( f \Gamma' \gamma' \vert \vec D. \, \vec {\cal E} 
\vert v_1 \Gamma_1 \gamma_1 ) ( v_1 \Gamma_1 \gamma_1
\vert \vec D. \, \vec {\cal E} 
\vert v_2 \Gamma_2 \gamma_2 ) \times \cr 
& \times ( v_2 \Gamma_2 \gamma_2 \vert H_3 \vert i \Gamma \gamma ) + 
 \hbox {term} \ [\vec D. \, \vec {\cal E} \vert H_3 \vert 
\vec D. \, \vec {\cal E}] + 
 \hbox {term} \ [H_3 \vert \vec D. \, \vec {\cal E} \vert 
\vec D. \, \vec {\cal E}] 
} 
$$
where the initial, intermediate and final state 
vectors are non-polluted. 

By following the same line of reasoning as in the case of 
intra-configurational transitions, we are left with the 
intensity formula$^{20)}$
$$
\eqalign{
S_{\Gamma \to \Gamma'} = {\hbox {Re}} [
\sum_{k, \ell = 0,2} & \sum_{r, s} \sum_{\Gamma''} 
I_1 [k \ell r s \Gamma'' ; \Gamma \Gamma'] \sum_{\gamma''} \left\{ 
{\cal E} \, {\cal E} \right\} ^{(k)}    _{r \Gamma'' \gamma''} \> 
\left( \left\{ 
{\cal E} \, {\cal E} \right\} ^{(\ell)} _{s \Gamma'' \gamma''}\right)^* 
\cr
+ \sum_{ k = 0,2} & \sum_{r, s} \sum_{\Gamma''} 
I_2 [k    2 r s \Gamma'' ; \Gamma \Gamma'] \sum_{\gamma''} \left\{ 
        {\cal E} \, {\cal E} \right\} ^{(k)} _{r \Gamma'' \gamma''} \> 
\left\{ {\cal E} \, {\cal E} \right\} ^{(2)} _{s \Gamma'' \gamma''} 
] 
}
\eqno (21)
$$
which parallels the formula (11). A detailed expression of the 
intensity parameters $I_1$ and $I_2$ will be found in the 
thesis by Daoud$^{20)}$ and in forthcoming papers.

\noindent {3.2. Illustration}

Let us consider the case of the configuration $4f$ in tetragonal symmetry 
with $G \equiv C_{4v}$ and examine the two-photon transitions
between the Stark levels of the shells $4f$ and $5d$
(i.e., $n \ell \equiv 4f$, $N \equiv 1$, $n' \ell' \equiv 5d$). 
There are four possible transitions since the initial and final 
states may have the symmetries $\Gamma_6$ and $\Gamma_7$. For a 
linear polarization, the application of the intensity formula 
(21) leads to 
$$
\eqalign{
S_{\Gamma_6 \to \Gamma_7} & = 
f \, \pi_2 + g \, \pi_3 + h \, \pi_4 + i \, \pi_5, 
\quad 
S_{\Gamma_7 \to \Gamma_7} = 
a' + b'\, \pi_1 + c'\, \pi_1^2 + d'\, \pi_2 + e'\, \pi_3 \cr
S_{\Gamma_7 \to \Gamma_6} & = 
f'\, \pi_2 + g'\, \pi_3 + h'\, \pi_4 + i'\, \pi_5, 
\quad
S_{\Gamma_6 \to \Gamma_6} = 
a  + b \, \pi_1 + c \, \pi_1^2 + d \, \pi_2 + e \, \pi_3 
} 
$$
where the angular functions $\pi_i$ ($i=1,2,3,4,5$) are defined by
$$
\pi_1 = 3 \cos^2 \theta - 1,            \
\pi_2 = \sin^2 2 \theta,                \
\pi_3 = \pi_2 \cos 2 \varphi,           \
\pi_4 = \sin^4 \theta \cos^2 2 \varphi, \
\pi_5 = \sin^4 \theta - \pi_4
$$
The various parameters $a, \cdots, i$ and $a', \cdots, i'$ 
are simple functions$^{20)}$ of the intensity parameters $I_1$ and 
$I_2$ occurring in (21).

\noindent {4. CLOSING REMARKS}

We have shown how $O(3) \supset G$ symmetry adaptation allows 
to derive intensity formulas for intra- and 
inter-configurational two-photon transitions for ions in 
molecular or solid-state environments. In particular, the 
number of independent parameters required for describing the 
polarization dependence of the transitions is determined by an 
ensemble of properties and rules which combine symmetry and 
physical considerations. The main results of this paper are 
formulas (11) and (21) for intra- and inter-configurational 
transitions, respectively. 

The polarization factors in (11) and (21) are under 
the control of the experimentalist. Both formulas depend on 
expansion coefficients
$c (\alpha S L J a \Gamma ; i)$ and $c (\alpha' S' L' J' a' \Gamma' ; f)$. 
These coefficients (model dependent) can be obtained by 
optimizing Hamiltonians, for the ion in its 
environment, involving at least Coulomb, spin-orbit
and crystal-field interactions~; the introduction of more sophisticated
interactions may be useful to take covalency effects into
account$^{2)}$. Alternatively, the expansion coefficients can be
considered as free parameters. Furthermore, in (11) and (21) we 
have reduced matrix elements (configuration dependent), 
isoscalar factors for $O(3) \supset G$ (group theory dependent), and 
$C[(k_Sk_L)k]$ parameters (mechanisms dependent). 
As a conclusion, there are three 
ways to deal with the intensity parameters in (11) and (21)~: 
they can be calculated from first principles, or considered as 
phenomenological parameters or determined in a mixed approach. 

The author is very grateful to the Organizing Committee of the 
School for inviting him to give this lecture. 

\noindent {REFERENCES}

\baselineskip 0.5 true cm



\item{1} 
Kibler, M., C.~R.~Acad.~Sc.~(Paris) B \souligne{268}, 1221 (1969)~; 
Kibler, M.R. and Guichon, P.A.M., Int.~J.~Quantum Chem.~\souligne{10}, 
87 (1976)~; 
Kibler, M.R. and Grenet, G.,      Int.~J.~Quantum Chem.~\souligne{11}, 
359 (1977)~; 
Kibler, M.R., in~: Recent Advances in Group 
Theory and Their Application to Spectroscopy, Ed.~J.C.~Donini 
(Plenum, 1979)~; Int.~J.~Quantum 
Chem.~\souligne{23}, 115 (1983)~; Croat.~Chem.~Acta 
\souligne{57}, 1075 (1984).
 
\item{2} Kibler, M. and Grenet, G., Report LYCEN 8656 (IPN Lyon, 1986).

\item{3} Axe, J.D., Jr., Phys.~Rev.~\souligne{136}, A42 (1964).

\item{4} Inoue, M. and Toyozawa, Y., J.~Phys.~Soc.~Japan 
\souligne{20}, 363 (1965). 

\item{5} Bader, T.R. and Gold, A., Phys.~Rev.~\souligne{171}, 
997 (1968).

\item{6} Apanasevich, P.A., Gintoft, R.I., Korolkov, V.S., 
Makhanek, A.G. and Skrip\-ko, G.A., Phys.~Status Solidi (b) 
\souligne{58}, 745 (1973)~; Makhanek, A.G. and Skrip\-ko, G.A., 
Phys.~Status Solidi (a) \souligne{53}, 243 (1979)~; 
Yuguryan, L.A., Preprints N$^\circ$ 232 and 233, Inst.~Fiz.~Akad.~Nauk 
BSSR, Minsk (1980)~; Makhanek, A.G., Korolkov, V.S. 
  and Yuguryan, L.A., Phys.~Status Solidi (b) \souligne{149}, 
231 (1988).

\item{7} Judd, B.R. and Pooler, D.R., J.~Phys.~C \souligne{15}, 
591 (1982).

\item{8} Downer, M.C. and Bivas, A., Phys.~Rev.~B 
\souligne{28}, 3677 (1983)~; Downer, M.C., in~: Laser 
Spectroscopy of Solids II, Ed.~W.M.~Yen (Springer, 1989).

\item{9} Reid, M.F. and Richardson, F.S., Phys.~Rev.~B 
\souligne{29}, 2830 (1984). 

\item{10} Sztucki, J. and Str\c ek, W., Phys. Rev. B 
\souligne{34}, 3120 (1986)~; Chem. Phys. Lett. \souligne{125}, 
520 (1986)~; Chem.~Phys.~\souligne{143}, 347 (1990).

\item{11} Jankowski, K. and Smentek-Mielczarek, L., 
Molec.~Phys.~\souligne{60}, 1211 (1987)~; Smentek-Mielczarek, L. 
and Hess, B.A., Jr., Phys.~Rev.~B \souligne{36}, 1811 (1987). 

\item{12} Kibler, M. and G\^acon, J.C., Croat.~Chem.~Acta 
\souligne{62}, 783 (1989). 

\item{13} G\^acon, J.C., Marcerou, J.F., Bouazaoui, M., 
Jacquier, B. 
and Kibler, M., Phys.~Rev.~B \souligne{40}, 2070 (1989)~; 
G\^acon, J.C., Jacquier, B., 
Marcerou, J.F., Bouazaoui, M. and Kibler, M., 
J.~Lumin.~\souligne{45}, 162 (1990)~; 
G\^acon, J.C., Bouazaoui, M., Jacquier, B., Kibler, M., 
Boatner, L.A. and Abraham, M.M., Eur.~J.~Solid State 
Inorg.~Chem.~\souligne{28}, 113 (1991). 

\item{14} Kibler, M., in~: Symmetry and Structural Properties 
of Condensed Matter, Eds.~W.~Florek, T.~Lulek and M.~Mucha 
(World, 1991).  

\item{15} Campochiaro, C., McClure, D.S., Rabinowitz. P. 
and Dougal, S., Phys.~Rev. B \souligne{43}, 14 (1991). 

\item{16} Kibler, M. and Daoud, M., Report LYCEN 9117 (IPN Lyon, 1991).

\item{17} Sztucki, J., Daoud, M. and Kibler, M., Phys.~Rev.~B (in press).

\item{18} Loudon, R., The Quantum Theory of Light 
(Clarendon, 1973).

\item{19} Cohen-Tannoudji, C., Dupont-Roc, J. and 
Grynberg, G., Processus d'inter\-action entre photons et atomes 
(InterEditions et Editions du CNRS, 1988).

\item{20} Daoud, M., th\`ese de Doctorat (Universit\'e 
Lyon-1, in preparation). 

\bye